\documentclass[useAMS,usenatbib]{mn2e}
\usepackage{graphicx}

\title[Polarization properties of Comet Halley]
{Modeling polarization properties of comet 1P/Halley using a mixture of compact and aggregate particles}
\author[H. S. Das, D. Paul, A. Suklabaidya,   and A. K. Sen]{H. S. Das\thanks{E-mail:
hsdas@iucaa.ernet.in (HSD)}, D. Paul, A. Suklabaidya,  and A. K. Sen\\
Department of Physics, Assam University, Silchar 788011, India.\\
}
\begin{document}

\date{Accepted xxxx. Received xxxx; in original form xxxx}

\pagerange{\pageref{firstpage}--\pageref{lastpage}} \pubyear{2011}

\maketitle

\label{firstpage}

\begin{abstract}
The \textit{in situ} measurement of Comet 1P/ Halley and the `Stardust' returned samples of comet Wild 2 showed the presence of a mixture of compact and aggregate particles, with composition of both silicates and organic refractory in cometary dust. Recently, the  result obtained from `Stardust' mission suggests that the overall ratio of compact to aggregate particles is 65:35 (or 13:7) for Comet 81P/Wild 2 (Burchell et al. 2008, Meteoritics \& Planetary Science, 43, 23). In the present work, we propose a model which considers cometary dust as a mixture of compact  and aggregate particles, with composition of silicate and organic. We consider compact particles as spheroidal particles and aggregates as ballistic cluster-cluster aggregate (BCCA) and ballistic agglomeration with two migrations (BAM2) aggregate with some size distribution. The mixing ratio of compact to aggregate particles is taken to be 13:7. For modeling Comet 1P/ Halley, the power-law size distribution $n(a) \sim a^{-2.6}$, obtained from re-analysis of the Giotto spacecraft data, for both compact and aggregate  particles is taken.  We take a mixture of BAM2 and BCCA aggregates with a lower cutoff size around 0.20$\mu m$ and an upper cutoff of about 1$\mu m$. We also take a mixture of prolate, spherical and oblate compact particles with  axial ratio (E) from 0.8 to 1.2 where a lower cutoff size around 0.1$\mu m$ and an upper cutoff of about 10$\mu m$ are taken.   Using T-matrix code for polydisperse spheroids (0.1$\mu m$ $\le a \le 10\mu m$) and Superposition T-matrix code for aggregates (0.2$\mu m$ $\le a_{v} \le 1\mu m$), the average simulated polarization curves are generated which can best fit the observed polarization data at the four wavelengths $\lambda$ = 0.365$\mu m$, 0.485$\mu m$, 0.670$\mu m$ and 0.684$\mu m$. The suitable mixing percentage of aggregates emerging out from the present modeling corresponds to 50\% BAM2 and 50\% BCCA particles and silicate to organic mixing percentage corresponds to 78\% silicate and 22\% organic in terms of volume.   The present model successfully reproduces the observed polarization data, especially the negative branch, for comet 1P/Halley at the above four wavelengths, more effectively as compared to other work done in the past. It is found that among the aggregates, the BAM2 aggregate plays a major role, in deciding the cross-over angle and depth of negative polarization branch.

\end{abstract}


\section{Introduction}
The study of polarization of the scattered radiation from comets,
over various scattering angles and wavelengths, gives valuable
information about the nature of cometary dust. The analysis of
polarization data gives information about the physical properties
of the cometary dust, which include size distribution, shape and
complex refractive indices.

The \emph{in situ} dust measurement of Comet 1P/Halley gave the first
direct evidence of grain mass distribution (Mazets et al. 1986).
Mukai, Mukai \& Kikuchi (1987) and Sen et al. (1991) analyzed the
polarization data of Comet 1P/Halley using power law dust
distribution (Mazets et al. 1986) and using Mie theory derived a
set of refractive indices of cometary grains. The dust
distribution function derived by Mazets et al. (1986) is actually
based on only Vega 2 results, while Lamy, Gr\"{u}n \& Perrin (1987) derived the
grain size distribution function for Comet 1P/Halley by comparing the
data from spacecrafts Vega 1, Vega 2 and Giotto. Much later
this dust distribution function was used by  Das, Sen
\& Kaul (2004) to analyze the polarization data of a number of comets including
Comet 1P/Halley.

Several investigators made useful polarimetric measurements of
Comet 1P/Halley through International 1P/Halley Watch (IHW) filters
(Bastien, Menard \& Nadeau 1986; Kikuchi et al. 1987; Le Borgne, Leroy \& Arnaud 1987;
Sen et al. 1991; Chernova, Kiselev \& Jockers  1993). The polarization data of
Comet 1P/Halley were analyzed by several investigators  using Mie
theory which assumes the dust particles to be spherical (Mukai et
al. 1987; Sen et al. 1991; Das et al. 2004).  However, the naturally occurring cometary grains cannot be ideal compact spheres, as required by Mie theory. The Mie theory was used, as it is more convenient and direct, with fewer numbers of free parameters required for modeling. Das \& Sen (2006)
studied the non-spherical dust grain characteristics of Comet Levy
1990XX using the T-matrix theory. They found that compact prolate
grains (with axial ratio = 0.486) as compared to spherical grains can better explain the
observed linear polarization data. Assuming an individual cometary grain to be an aggregate of several monomers, Das et al. (2008a)  again analyzed the observed polarization data of Comet C/1990 K1 Levy and successfully reproduced
the polarization curve through simulations, where the fit was still better. The  $\chi ^2_{\textrm{min}}$ value for the aggregates was found to be 4.2 whereas the value obtained by Das \& Sen (2006) for compact prolate grains was 5.22. Thus it was concluded that aggregate
particles can produce a still better fit to the observed data as compared to compact prolate grains.
Again, Das et al. (2008b) successfully explained the
polarization characteristics of comet C/1995 O1 Hale-Bopp at $\lambda = 0.485$ $\mu m$ and 0.684 $\mu m$ using aggregate dust model.
However aggregate dust model used in the previous work was restricted to single size of monomer with same size parameter at different wavelengths. More recently, Das et al. (2010) included the size distribution for aggregates and studied the observed polarization data of comet C/1996 B2 Hyakutake at $\lambda = 0.365$$\mu m$, 0.485$\mu m$ and 0.684$\mu m$.

It is now well accepted from \textit{in situ} measurement of comets and `Stardust' returned samples of comet Wild 2 that cometary dust consists of a mixture of compact particles and  aggregates (Lamy et al. 1987; Fomenkova et al. 1999;  H\"{o}rz et al.  2006; Zolensky et al. 2006, Burchell et al. 2008 etc.). Lasue et al. (2009)
studied comet 1P/Halley and comet C/1995 O1 Hale-Bopp using a mixture of fluffy
aggregates and compact solid grains. They developed a model of light scattering by a size distribution of  aggregates of up to 256 submicron-sized grains (spherical or spheroidal) mixed with single spheroidal particles. A good fit of the positive polarization observations of 1P/Halley
had been obtained by them with a power law size distribution ($a^{-2.8}$ with a
lower cutoff of 0.26 $\mu m$ and an upper cutoff of 38 $\mu m$) with
a mixture of silicates (between 40\% and 67\% in volume) and
more absorbing organic material (between 33\% and 60\% in
volume). The fits deduced from
their model show that the negative polarization branch is not deep enough to match the observed polarization data,
especially for comet 1P/Halley. Although the fits are found to be good for the positive part
of the polarization. Recently, Kolokolova \& Kimura (2010) modeled cometary dust as a mixture of compact  particles (made of silicate) and aggregates (made mainly of organics and 1P/Halley like composition). Using a size distribution function $a^{-3}$ for compact particles and 256 number of BCCA aggregates, they reproduced the polarimetric data, including negative polarization at small phase angles  and the positive polarization with the maximum value less than 30\% at the phase angle around $90^{\circ}$ and red polarimetric color. However, their model reproduced feature common to `dusty comets' polarization curves but they did not use a chi square fitting procedure to compare with observed data for a given comet.

In the present work, a model for cometary dust with a mixture of compact spheroidal particles and aggregates with size distribution are proposed to study the observed polarization data of Comet 1P/Halley at $\lambda = 0.365$$\mu m$,
0.485$\mu m$, 0.670$\mu m$ and 0.684$\mu m$.

\section{Dust model}
The \textit{in situ} measurement of comet 1P/Halley and the `Stardust' returned samples of comet Wild 2 showed the presence of a mixture of compact and aggregate particles with composition of silicates and organic refractory. Moreno et al. (2007) conducted a systematic approach to test whether a collection of compact particles can reproduce the observed properties of cometary dust. Using a model of spheroidal particles, they found that the axial ratio should be either E = 2 (oblate) or E = 0.5 (prolate). The refractive indices lie within a range $n$ = 1.6 -- 1.7 and $k$ = 0.05 -- 0.1. They also studied a more complex model based on size distributions of irregularly shaped particles composed by a varying number of cubes as elementary units. The models considering irregularly shaped and compact particles with different structures showed results close to the observations. However, the weakness of their model of compact structures was that the maxima in the linear polarization values did not take place in the $90^{\circ} - 100^{\circ}$ phase angle region as observed. Recently, Kolokolva \& Kimura (2010) modeled cometary dust as a mixture of compact spheroidal and aggregate particles. The compact particles which they considered to be a mixture of oblate and prolate spheroids with axial ratio within the range 1 -- 2.5 and aggregates were taken to be Ballistic Cluster-Cluster Aggregate (BCCA).

In the present work, we propose a model which considers cometary dust as a mixture of compact and aggregate  particles. Since the \textit{in situ} analysis of dust samples exhibits the overall ratio of compact to aggregate particles to be 65:35 (Burchell et al. 2008), so we take the same value in our analysis. For modeling comet 1P/Halley, we will use a power-law size
distribution, $n(r) = dn/da \sim a^{-2.6}$ for both compact particles and aggregates, obtained from a re-analysis
of the Giotto data by Fulle et al. (2000).

We consider compact spheroidal particles with a size distribution from 0.1$\mu m$ to 10$\mu m$. The particles are presented by multishaped, polydisperse mixture of spheroids. We consider a mixture of prolate, spherical and oblate compact particles with  axial ratio (E) from 0.8 to 1.2. Computations of light scattering by plain and coated particles are made through codes adapted from T-matrix code (Mishchenko \& Travis, 1996).

We build the aggregates using ballistic
aggregation procedure (Meakin 1983, 1984). Two different
models of cluster growth are taken: first via single-particle aggregation
and then through cluster-cluster aggregation. These aggregates are
built by random hitting and sticking particles together. The first
one is called Ballistic Particle-Cluster Aggregate (BPCA) when the
procedure allows only single particles to join the cluster of
particles. If the procedure allows clusters of particles to stick
together, the aggregate is called Ballistic Cluster-Cluster
Aggregate (BCCA). Actually, the BPCA clusters are more compact
than BCCA clusters (Mukai et al. 1992). The porosity of BPCA and
BCCA particles of 128 monomers has the values 0.90 and 0.94,
respectively and the fractal dimension of BPCA and BCCA is $D \approx 3$ and 2, respectively. A systematic explanation on dust aggregate model is
already discussed in our previous work (Das et al. 2008a).

Recently, Shen et al. (2008) considered three different classes of clusters
distinguished by aggregation rules. These are BA ("ballistic agglomeration"), BAM1 ("ballistic agglomeration
with one migration") and BAM2 ("ballistic agglomeration with two migrations"). They developed a set of
parameters to characterize the irregular structure of these aggregates. Actually BA cluster is identical
with BPCA cluster. The geometry of BAM1 and BAM2 clusters are random but less porous than BA clusters. The
effective porosity (P) increases from BAM2 $\rightarrow$ BAM1 $\rightarrow$ BA. The porosity of BAM2 structure having 64 number of monomers have the value $P \approx 0.5$ and the fractal dimension is $D \approx 3$ (Shen et al. 2008). The aggregates are taken from web 1 (see reference).

In our model, we take  same cloud of particles, i.e., same type of particles (compact spheroidal  and porous (BCCA + BAM2) particles) and same size distribution ($n(r) = dn/da \sim a^{-2.6}$), to fit the observed data at all wavelengths.

In our simulation, we divide the present work into two phases:

\begin{enumerate}
\item We first take BCCA aggregates and then mix with compact spheroidal particles with 65:35 mixing ratio. The result obtained from  this modeling will be discussed in \textit{Section 4}.

  \item We then consider more compact aggregate BAM2 (having  porosity (P) $\sim$ 0.50 approximately) which is mixed with highly porous BCCA clusters (P $\sim$ 0.9) with some variable mixing ratio ($\beta$). Then the aggregate mixture is  mixed with compact particles with 65:35 mixing ratio. Here we take composition of both silicate and organic with variable mixing ratio ($\gamma$).
\end{enumerate}

The free parameters used in the model are as follows:
\begin{itemize}
  \item the mixing ratio ($\beta$) between BCCA and BAM2.
  \item the mixing ratio ($\gamma$) between silicate and organic.
\end{itemize}

We use $\chi ^2$ - minimization technique to evaluate the best fit values of the above free parameters by the following equation:

\begin{equation}
\chi_{\textrm{pol}}^2 = \sum_{i=1}^N  \left|
\begin{array}{c}
\frac{\textrm{P}_{obs} (\theta_i,\lambda) -
\textrm{P}_{theo}(\theta_i,\lambda)} {
 \textrm{E}_p (\theta_i,\lambda)}
\end{array}
\right|^2
\end{equation}

Here, $\textrm{P}_{obs} (\theta_i,\lambda)$ is the degree of
linear polarization observed at scattering angle $\theta_i$ (i =
1,2,....,N) and wavelength $\lambda$,
$\textrm{P}_{theo} (\theta_i,\lambda)$ is the polarization values
obtained from model calculations and $\textrm{E}_p
(\theta_i,\lambda)$ is the error in the observed polarization at
scattering angle $\theta_i$ and wavelength ($\lambda$).
It is also observed that this technique of minimization of $\chi
^2$ is quite unique. The value of $\chi_{\textrm{min}}^2$ gives
the confidence level on our best fit values of  $\beta$ and $\gamma$ and also in
the overall fitting procedure. Some preliminary work on combined dust model has been already reported in Das \& Sen (2011).

\section{Composition}
The \emph{in situ} observation of comets, laboratory analysis of samples of
IDP and remote infrared spectroscopic study of comets give useful information
about the composition of cometary dust. The \emph{in situ}
measurement, of impact-ionization mass spectra of Comet 1P/Halley's
dust, has suggested that the dust consists of magnesium-rich
silicates, carbonaceous materials, and iron-bearing sulfides
(Jessberger et al. 1988; Jessberger 1999). These materials are
also known to be the major constituents of IDPs (Brownlee et al.
1980). The studies of comets and IDPs have
shown the presence of amorphous and crystalline silicate minerals
(e.g.  forsterite, enstatite) and organic materials (Hanner \&
Bradley 2004). Laboratory studies have  shown that majority of the
collected IDPs fall into one of the three spectral classes. These
observed profiles indicate the presence of  pyroxene, olivine and
layer lattice silicates. This is in good agreement with results
obtained from Giotto and Vega mass spectrometer observations of
Comet 1P/Halley (Lamy et al. 1987). The infrared (IR) measurement of
comets has also provided important information on the silicate
compositions in cometary dust. The spectroscopic studies of
silicates have shown the predominance of both crystalline and
amorphous silicates consisting of pyroxene or olivine grains
(Wooden et al. 1999; Hayward, Hanner \& Sekanina 2000, Bockel\'{e}e - Morvan et
al. 2002 etc.). Mg-rich crystals are also found within IDPs and
are predicted by comparing the IR spectral features of Comet
C/1995 O1 Hale-Bopp with synthetic spectra obtained from laboratory studies
(Hanner  1999; Wooden et al. 1999, 2000).  `Stardust' samples have
also confirmed a variety of olivine and pyroxene silicates in
Comet 81P/Wild 2 (Zolensky et al. 2006).

It is to be noted that though we used only two free parameters $\beta$ and $\gamma$ in our model, the refractive indices of silicate and organic can be used as other free parameters. However, we have limited ourselves to the value taken from standard references, because with many free parameters the computational time becomes very long.  In our computation, we take the refractive indices of silicate (especially amorphous pyroxene)  from Dorschner et al. (1995). The refractive indices of the amorphous pyroxene (Mg$_{x}$Fe$_{1-x}$SiO$_3$,
where $x$ is the Mg number, $x$ = $\frac{\textrm{Mg}}{\textrm{Mg}+\textrm{Fe}}$, $x$= 0.4, 0.5, 0.6, 0.7, 0.8, 0.95 and 1.0) are  reported by them for different values of $x$. We select $x = 0.5$ to consider the equal number of Mg and Fe in the pyroxene formula. However, we do not claim that the choice of $x = 0.5$ is unique. The values are calculated by linearly interpolating the data obtained from laboratory studies. The refractive indices are given by (1.722,0.101) at 0.365$\mu m$, (1.692,0.0492) at 0.485$\mu m$, (1.673,0.0198) at 0.670$\mu m$ and (1.672,0.0185) at 0.684$\mu m$.  The refractive indices of organic are taken from Jenniskens (1993) and the values  are given by (1.679,0.536) at 0.365$\mu m$, (1.842,0.459) at 0.485$\mu m$, (1.942,0.357) at 0.684$\mu m$ and (1.949,0.349) at 0.684$\mu m$. The refractive indices of silicate and organic at 0.485$\mu m$ have been already used by Das \& Sen (2011) to model the optical polarization of comets.

\section{Numerical simulation}
For modeling comet 1P/Halley, we use a power-law size
distribution, $n(a) = dn/da \sim a^{-2.6}$ for both compact particles and aggregates, obtained from a re-analysis
of the Giotto data by Fulle et al. (2000). The observed linear polarization data of Comet
1P/Halley is taken from Bastien et al. (1986), Gural'Chuk et al. (1987), Kikuchi et al. (1987),
Le Borgne et al. (1987), Sen et al. (1991) and Chernova et al.
(1993) at $\lambda = 0.365$$\mu m$, 0.485$\mu m$, 0.670$\mu m$ and 0.684$\mu m$.

We calculate the scattering properties of spheroidal compact particles using T-matrix code (Mishchenko \& Travis, 1996) for $0.1\le a \le 10\mu m$, where $a$ is the equal volume sphere radius of the particle. The step size used to integrate the size distribution is 0.01$\mu m$. We also calculate the scattering properties of the BCCA and BAM2
clusters using superposition \textsc{t-matrix} code, which gives
rigorous solutions for ensembles of spheres (Mackowski \&
Mishchenko 1996).

The size of the individual monomer in a cluster plays an important
role in scattering calculations. These have been confirmed by the
results of previous work on dust aggregate model (Kimura et al. 2006; Petrova et al. 2004; Hadamcik et al. 2006; Bertini et
al. 2007; Das et al. 2008a). The radius of an aggregate
particle can be described by the radius
of a sphere of equal volume given by $a_v = a_m N^{1/3}$, where N is the number of monomers
in the aggregate. In the present work, BCCA with 128 monomers and BAM2 with 64 monomers are taken. As we had computational limitation and BAM2 cluster takes longer computational time compared to BCCA particles, we had to restrict ourselves the number of monomers to 64 only for BAM2 particles. In our calculation, averages of three random realizations are taken for both BCCA and BAM2.
The size range of the monomer is taken in the range  $0.05\mu m \le a_m \le 0.20 \mu m$. Thus the lower cutoff radius of the cluster is 0.2 $\mu m$ and the upper cutoff is 1 $\mu m$.  It is to be noted that since the number of monomers is fixed in each type of aggregate, the distribution in monomer sizes is essentially the size distribution of aggregates. For a particular type of aggregate with fixed N, the size distribution is just $dn/da_v \sim a_v^{-2.6}$. The step size used to integrate the size distribution is 0.01$\mu m$.

We start calculation considering only BCCA particles and then mix with compact spheroidal particles with 13:7 mixing ratio. It has been checked that the mixing of compact spheroidal grains and aggregates will not help much in producing deeper negative polarization branch beyond \textbf157$^{\circ}$.  It has been also observed from Lasue et al. (2009) that the fits deduced from  their modeling do not show deep negative polarization branch for comet 1P/Halley.

Using aggregate dust models with BAM2 geometry and moderate porosity (P$\approx$0.6), Shen et al. (2009) reproduced albedo and polarization for cometary dust, including negative polarization observed at scattering angles beyond 160$^{\circ}$. To study the effect of BAM2 structure, we now start computation at $\lambda = 0.485$$\mu m$ with BCCA and BAM2 particles with different mixing ratio ($\beta$) and then finally mix with compact spheroidal particles having mixing ratio 65:35 (or 13:7), where $\gamma$ is taken to be 3:1. In \textbf{Fig.1}, the polarization curves are generated for $\beta$ = 1:3 and 1:1, which actually correspond to (25\% BCCA + 75\% BAM2) and (50\% BCCA + 50\% BAM2) particles for a size distribution $n(a) \sim a^{-2.6}$. The size range for the aggregates (BCCA and BAM2) and compact spheroidal particles
 is taken to be $0.2 \le a \le 1.0\mu m$ and $0.1 \le a \le 10\mu m$ respectively. We also generate the polarization curves separately with BCCA and BAM2 particles. \textbf{Fig.1(a)} shows the average polarization curve obtained from the mixing of compact spheroidal particles and aggregates (BCCA and BAM2). The mixing ratio between BCCA and BAM2 is 1:3. In \textbf{Fig.1(b)}, the curve 1 corresponds to average polarization curve in the range $150^0-180^0$ obtained from the mixing of compact and BAM2 particles only, curve 2 with $\beta = 1:3$, curve 3 with $\beta = 1:1$ and curve 4 obtained from the mixing of compact spheroidal particles and BCCA particles only.

It is clear from \textbf{Fig. 1} that the existence of BAM2 structure (which is more compact than BCCA) becomes crucial in producing the deeper negative polarization branch.  Thus the introduction of BAM2 aggregate in the aggregate mixture will help to reproduce the negative polarization well which was not possible in previous study by several investigators for comet 1P/Halley. Actually, interplanetary dust particles (IDP)  may contain both porous and compact aggregates. So it will be more realistic if we consider aggregates to be a mixture of more compact BAM2 (P $\sim$ 0.5) and more porous BCCA (P $\sim$ 0.9) clusters with some mixing ratio $\beta$.

We now use $\chi ^2$ minimization technique to evaluate the best-fitting
values of  $\beta$ and $\gamma$ which can fit to the observed
polarization data. We have already used this minimization
technique to fit the observed linear polarization data of some comets
(Das et al. 2008a,b; Das et al. 2010 and Paul et al. 2010), with aggregate
models of dust. We need to fine-tune the free parameters $\beta$ and $\gamma$ in the model to
make the best fit to the observed linear polarization data of Comet
1P/Halley. Some preliminary work on combined dust model has been reported by  Das \& Sen (2011) where they used the same technique to simulate the observed polarization data of comet Halley at 0.485$\mu m$. However their work is limited to single wavelength only.

The best fit values of $\beta$ and $\gamma$ are found to be 1:1 and 78:22 at $\lambda = 0.365$$\mu m$,
0.485$\mu m$, 0.670$\mu m$ and 0.684$\mu m$.  The $\chi ^2_{\textrm{min}}$ values emerging out from the present analysis are 14.8, 47.2 and 32.5 at $\lambda = 0.365$$\mu m$, 0.670$\mu m$ and 0.684$\mu m$ respectively, whereas the value obtained by Das \& Sen (2011) for $\lambda = 0.485\mu m$ is  56.7.   The best-fitting average polarization curves at four wavelengths are shown in \textbf{Fig. 2}, \textbf{Fig. 3}, \textbf{Fig. 4} and \textbf{Fig. 5}.

\begin{figure}
\includegraphics[width=84mm]{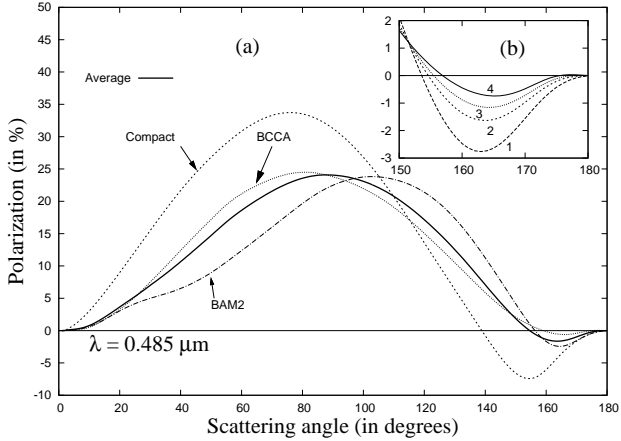}
\caption{ (a)The average polarization curve obtained from the mixture of compact spheroidal particles and aggregates (BCCA and BAM2) for a size distribution $n(a) \sim a^{-2.6}$  at $\lambda = 0.485$ $\mu m$ with $\beta = 1:3$, where the mixing ratio between compact and aggregates is taken to be 13:7 and $\gamma = 3:1$. (b) the curve 1 corresponds to average polarization curve in the range $150^0-180^0$ obtained from the mixing of compact and BAM2 particles only, curve 2 with $\beta = 1:3$, curve 3 with $\beta = 1:1$ and curve 4 obtained from the mixing of compact spheroidal particles and BCCA particles only.
 }
\end{figure}


\begin{figure}
\includegraphics[width=84mm]{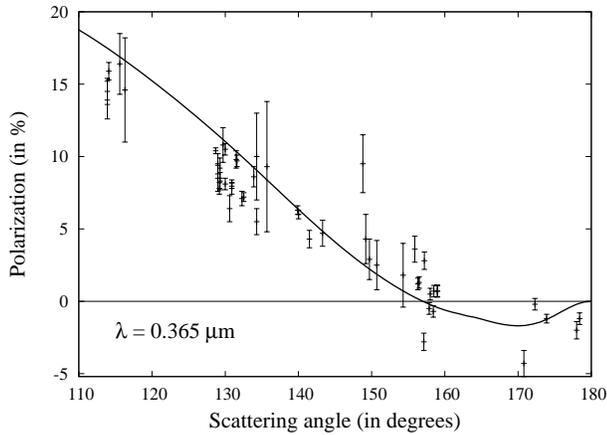}
\caption{Polarization values as observed at wavelength $\lambda = 0.365$ $\mu m$
 for Comet 1P/Halley by Bastien et al. (1986), Gural'Chuk et al.(1987), Kikuchi et al. (1987), Le Borgne et al.
 (1987), Sen et al. (1991) and Chernova et al. (1993). The solid  curve represents
 the best-fitting average polarization curve
 obtained  for compact particles and aggregates (BCCA and BAM2) for a size distribution $n(a) \sim a^{-2.6}$ at
 $\lambda = 0.365$ $\mu m$.}
\end{figure}

\begin{figure}
\includegraphics[width=84mm]{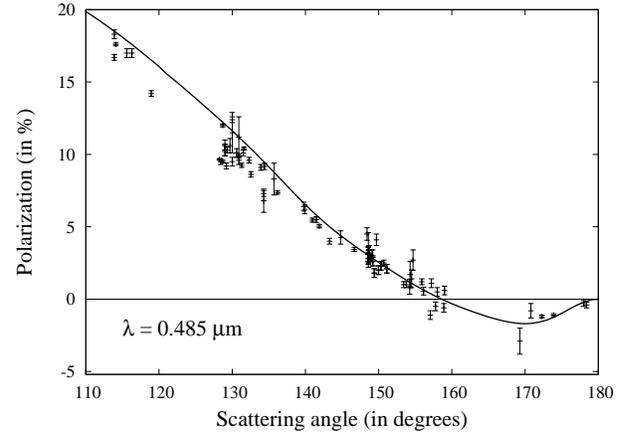}
\caption{The solid  curve represents
 the best-fitting average polarization curve
 obtained  for compact particles and aggregates (BCCA and BAM2) for a size distribution $n(a) \sim a^{-2.6}$ at
 $\lambda = 0.485$ $\mu m$, taken from Das \& Sen (2011).}
\end{figure}

\begin{figure}
\includegraphics[width=84mm]{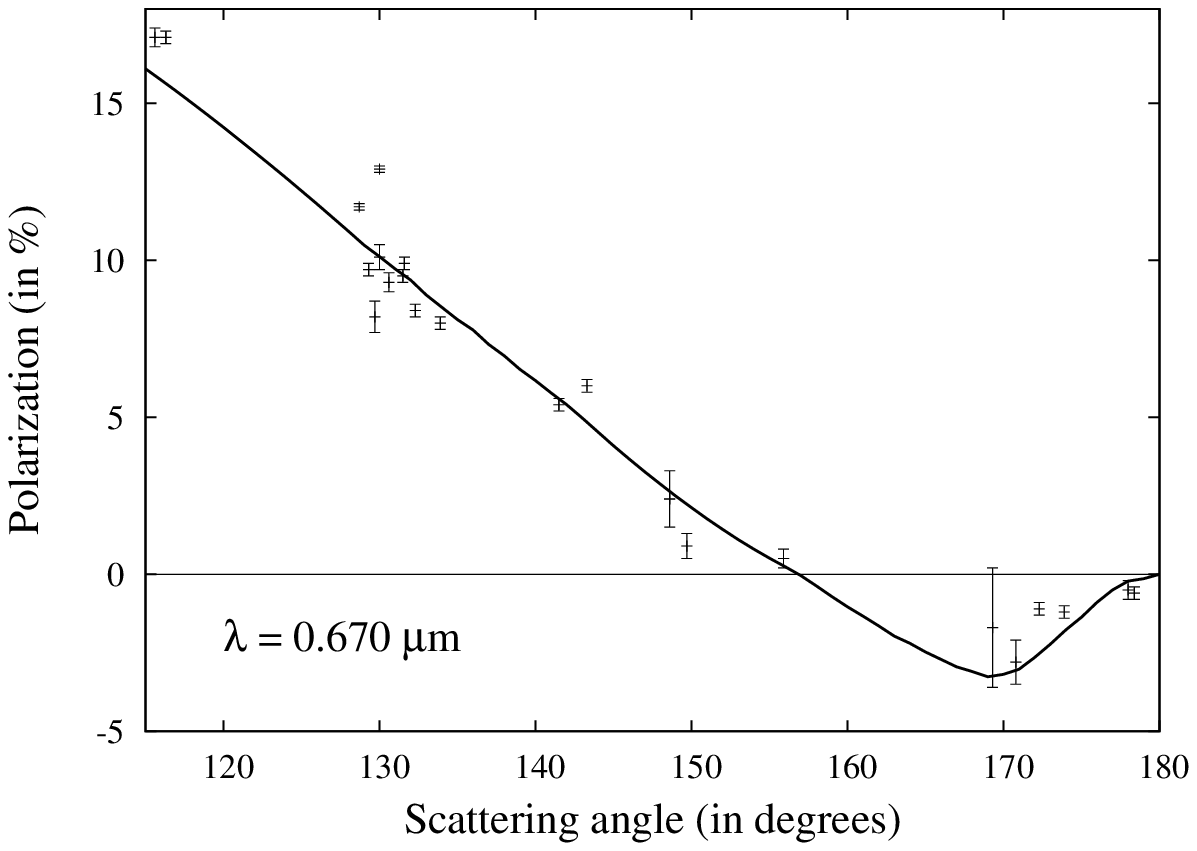}
\caption{The solid  curve represents
 the best-fitting average polarization curve
 obtained  for compact particles and aggregates (BCCA and BAM2) for a size distribution $n(a) \sim a^{-2.6}$ at
 $\lambda = 0.670$ $\mu m$.}
\end{figure}


\begin{figure}
\includegraphics[width=84mm]{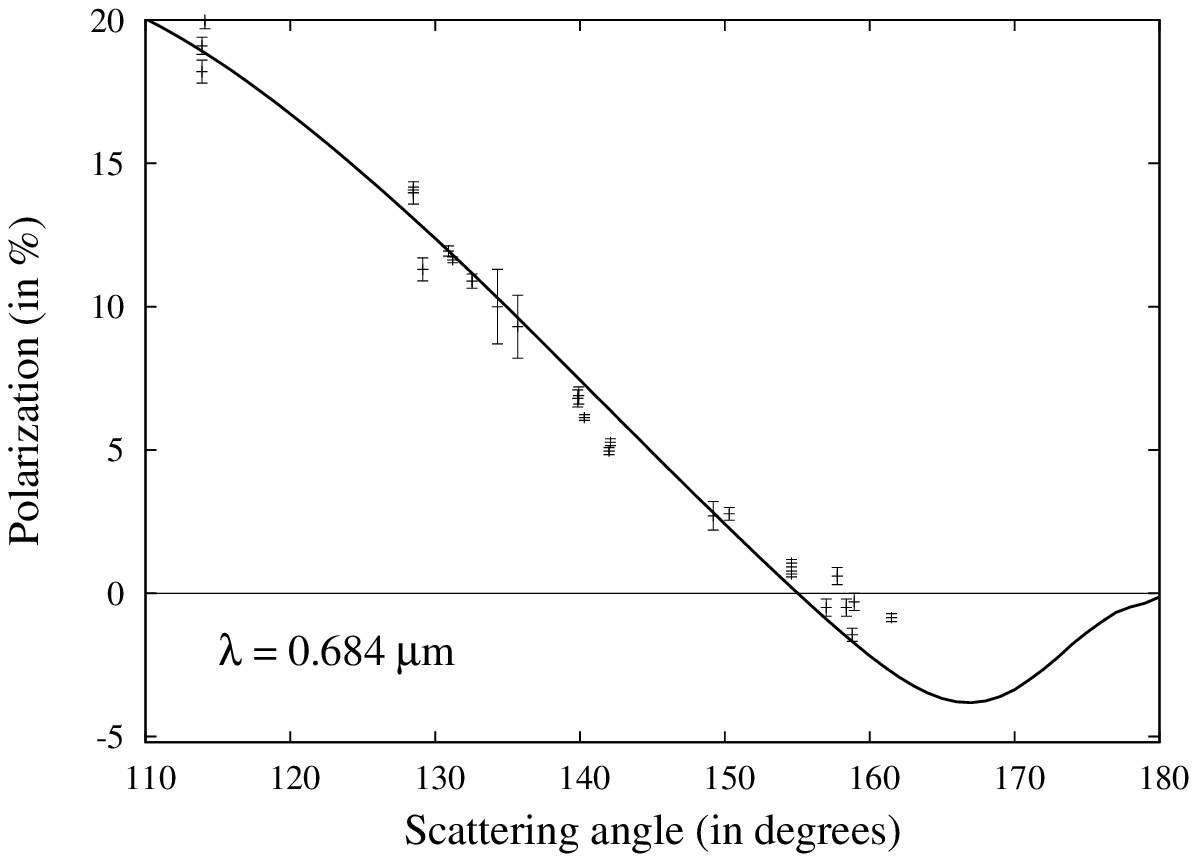}
\caption{The solid  curve represents
 the best-fitting average polarization curve
 obtained  for compact particles and aggregates (BCCA and BAM2) for a size distribution $n(a) \sim a^{-2.6}$ at
 $\lambda = 0.684$ $\mu m$.}
\end{figure}

\section{Discussion}
The \emph{in situ} measurement, of impact-ionization mass spectra of Comet 1P/Halley's
dust, has suggested that the dust consists of magnesium-rich
silicates, carbonaceous materials, and iron-bearing sulfides
(Jessberger et al. 1988; Jessberger 1999).  In our modeling we consider cometary dust as a mixture of compact and porous particles with composition of silicates and organic refractory. The silicate to organic ratio coming out from our present work is 39:11 or 78\% silicate and 22\% organic in volume. Thus it can be concluded that the silicate composition is dominating in comet 1P/Halley as compared to organic refractory.

The negative polarization feature of comet is one of the important feature observed in comets.
Many comets show negative polarization beyond $157^0$ ((Kikuchi et al. 1987; Chernova et al. 1993;
Ganesh et al. 1998 etc.). Several investigators (Greenberg \&
Hage 1990; Muinonen et al. 1996, 2007; Tishkovets et al. 2004; Petrova et al. 2004; Hadamcik et
al. 2007 etc.) have discussed the cause of negative
polarization in comets. Actually, it is important to fit the observed polarization data in the positive part as well as in the negative branch.
Using aggregate dust model, Das et al. (2008a,b; Paul et al. 2010) successfully reproduced the polarization curves including negative branch observed for comets C/1990 K1 Levy, C/1995 O1 Hale-Bopp, C/1996 B2 Hyakutake and C/2001 Q4 NEAT.  But it is now well accepted that cometary dust consists of compact and porous particles. Several investigators studied comet using a mixture of highly porous aggregates and compact solid grains. It has been observed that the plots are showing good fit to the positive part of the polarization, but do not show deeper negative polarization branch beyond $157^0$.

In our present work, we take a mixture of aggregates (highly and moderately porous) and then mix with compact spheroidal grains with 13:7 ratio.
It can be noticed from \textbf{Fig. 2}, \textbf{Fig. 3}, \textbf{Fig. 4} and \textbf{Fig. 5} that our modeling can successfully reproduce the positive part as well as the negative branch of the polarization at three different wavelengths. However, if we just withdraw the BAM2 structure from our model, the negative polarization branch will not be reproduced at proper scattering angle values. So it appears that the existence of BAM2 particles (which is more compact than BCCA) is very important in our grain model as it can reproduce a deeper negative polarization branch. Thus our modeling can help to explain the polarization characteristics of comet 1P/Halley successfully at different wavelengths.

The angular dependence of brightness and linear polarization of compact and porous clusters have been investigated by Tishkovets et al. (2004). They found that porous clusters are brighter almost in the whole angular range due to the larger cross section, and these clusters produce smoother polarization curves with higher maximum. The negative branch in the backscattering direction is shallower, because the wave interference and near-field effects are weaker within the aggregates. However, in compact clusters,   both the interference and the near-field effects play a major role in producing negative polarization branch. The negative branch is deeper for compact clusters as compared to porous clusters. The minimum is deeper, and inverted angle is shifted to smaller scattering angles. The negative polarization is mostly generated by the particles below the surface layer of the cluster, where the radiation field is inhomogeneous, and the amplitude, phase, and propagation direction of the wave change randomly (Tishkovets et al. 2004). It has been demonstrated by Petrova et al. (2004) that the external layer of the clusters plays important role in forming the polarization phase curve. The appearance of the negative polarization branch and its shape strongly depend on the sizes of the scattering elements and on the structure of the particle ensemble. In a subsequent work, Shen et al. (2009) studied the phase curve and polarization values, as produced by BAM1 and BAM2 and it was found that more compact BAM2 cluster show deeper negative polarization branch as compared to BAM1 and BCCA.

Before we conclude we may note that the $\chi^2$ values reported earlier  for comet C/1990 K1 Levy,  showed improvement of fit in Das et al. (2008a) with aggregate grains ($\chi^2$ value 4.2) as compared to Das \& Sen (2006) with compact prolate grains ($\chi^2$ value 5.22).

It is true that, these two $\chi^2$ values for comet C/1990 K1 Levy, are much lower than the $\chi^2$ value we obtained in the   present work for comet Halley.  It may be noted here that in the present work, the fit was made on the data points of comet Halley collected from various sources as observed by different groups of observers (for example at 0.485 micron, we have  86 data points collected from six different groups of observers).  Such data points collected from diverse groups of observers, will always have some inherent scatter in their values, as observations are made with different aperture sizes and with different sets of filters (with different central wavelengths and FWHM).  Besides different groups of observers use different instruments, with different spectral responses. When we club such data points from various sources, ideally one should calibrate all the observed data points to take into account the above effects. But it is a tedious job and normally such corrections are never made as in the present case.

On the other hand, in the two earlier work (Das \& Sen 2006 and Das et al. 2008a) on comet C/1990 K1 Levy, the fit was made on data values collected from a single source, viz Chernova et al. (1993).  Therefore, it is quite natural to expect that, the $\chi^2$ value in the present work on Halley will be higher than what has been obtained earlier for C/1990 K1 Levy. And this is due to the diversities in the sources of data points for Halley. For example at $\lambda$= 0.670$\micron$, if we exclude the  data point (position angle, polarization)=  (130.0, 12.9), the $\chi^2$ value just drops from 47.2 to 15.6.

For comet C/1990 K1 Levy the data points were only 16 as compared to 86 for Halley at wavelength $0.485 \micron$. Also for comet Halley we considered a much wider range of phase angle values as compared to Levy, which constrained our grain model further and increased the $\chi^2$  value.
What is important here to note that, for comet Halley no other grain model can generate a lower $\chi^2$ value (indicating a better fit)  than what  has been  reported by us in the present work.

\section{Conclusion}
\begin{enumerate}
\item A mixture of compact spheroidal grains and aggregates successfully explains the observed polarization data of comet 1P/Halley at at $\lambda = 0.365$$\mu m$, 0.485$\mu m$, 0.670$\mu m$ and 0.684$\mu m$.

  \item The positive part as well as the negative polarization have been successfully generated using the proposed combined model of cometary dust.

  \item With the introduction of  distribution of monomer sizes  and BAM2 cluster (more compact than BCCA), one can fit the observed polarization data much better, as compared to the previous work on comet 1P/Halley. It is also observed that existence of BAM2 structure becomes important in reproducing the deeper negative polarization branch.

  \item The best-fitting mixing ratio between BCCA and BAM2 ($\beta$) is found to be 1:1 (or 50\% BAM2 + 50\% BCCA).  Thus it can be concluded that porous grains in comet 1P/Halley are composed of both highly and moderate porous particles.

   \item The best-fitting mixing ratio between silicate and organic particles ($\gamma$) is found to be 39:11 (or 78\% silicate and 22\% organic in volume).
\end{enumerate}

\section{Acknowledgements}
We would like to thank anonymous referee for his/her helpful suggestions.  We acknowledge T. Mukai and Y. Okada for
help on the execution of BPCA and BCCA codes. We are also
thankful to M. Mishchenko et al., who made their superposition T-matrix code and T-matrix code for
randomly oriented spheroids publicly available. The authors HSD and AKS  acknowledge  Inter University Centre for
Astronomy and Astrophysics (IUCAA), Pune for its associateship
programme.

\end{document}